\documentclass[prl,twocolumn,showpacs]{revtex4}
\newcommand{\UQ}{ARC Centre of Excellence for Quantum-Atom Optics, 
School of Physical Sciences, University of Queensland, Brisbane, 
QLD 4072, Australia.}
\newcommand{\Virgin}{Department of Physics, University of Virginia, 382 McCormick Road, Charlottesville, Virginia 22904-4714, USA.} 
\usepackage{graphicx,psfrag}
\usepackage{bm}

\newcommand{\dfdx}[2]{\frac{\partial#1}{\partial#2}}

\begin{document}
\title{Bright tripartite entanglement in triply concurrent parametric oscillation}
\author{A.~S. Bradley}
\author{M.~K. Olsen}
\affiliation{\UQ}
\author{O. Pfister}
\author{R.~C. Pooser}
\affiliation{\Virgin}
\date{\today}
\begin{abstract}
We show that a novel optical parametric oscillator, based on three concurrent $\chi^{(2)}$ nonlinearities, can produce, above threshold, bright output beams of macroscopic intensities which exhibit strong tripartite continuous-variable entanglement. We also show that there are two ways that the system can exhibit a three-mode form of the Einstein-Podolsky-Rosen paradox, and calculate the extra-cavity fluctuation spectra that may be measured to verify our predictions.
\end{abstract}
\maketitle
Entanglement is a central property of quantum mechanics. In particular, continuous-variable (CV) entanglement is at the core of the Einstein-Podolsky-Rosen (EPR) paradox \cite{Einstein1935} and is also important for quantum teleportation, quantum information, and quantum cryptography~\cite{Braunstein,Braunstein2004}. Bipartite entanglement is now readily producible experimentally, and there has been some progress in the production of tripartite entangled beams, where entanglement was obtained by mixing squeezed vacua with linear optical elements~\cite{vanLoock2000,Aoki2003}. In our proposal, strong entanglement is created in the nonlinear interaction itself and, as we will demonstrate, is present above the operating threshold, where the entangled outputs are macroscopically bright beams. We confirm that the output fields satisfy criteria for measurable CV tripartite entanglement~\cite{vanLoock2003} and derive two types of  experimental EPR~\cite{Einstein1935} criteria which are applicable to this system. We then show analytically and numerically that the proposed system demonstrates these properties, both above and below threshold. Note that the tripartite CV entangled state thus created tends towards a GHZ state in the limit of infinite squeezing, but is analogous to a W state for finite squeezing \cite{Braunstein2}.
\par
\begin{figure}[!htb]
\includegraphics[width=1\columnwidth]{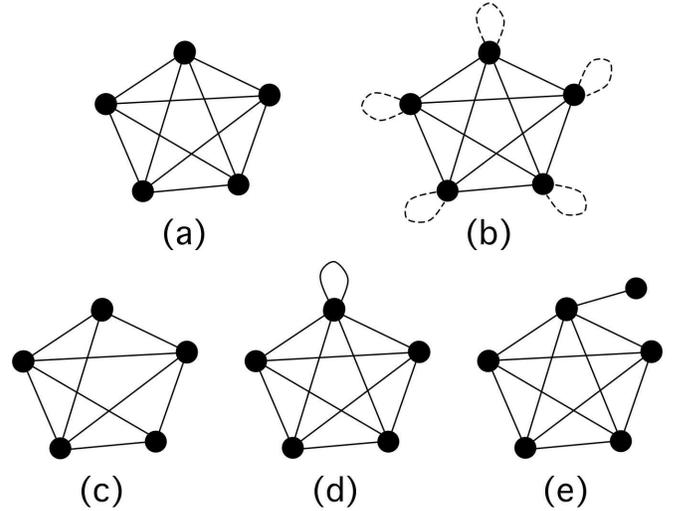}
\caption{Graphical representation of multipartite CV entanglers for $N=5$. Each dot represents a cavity field mode and the lines represent their respective nonlinear couplings. The solid lines in a graph represent, indifferently, all down- or all up-converting nonlinear interactions; the dashed lines then represent the interactions opposite to the solid line ones, i.e.\ all up- or down-converting, respectively. (a): ideal concurrence, yields maximal (GHZ) entanglement in the limit of infinite squeezing. (b): van Loock and Braunstein's ideal case: yields maximal (GHZ) entanglement in the limit of infinite squeezing. (c): nonideal concurrence due to an incomplete set of interactions. (d): nonideal concurrence due to an undesirable degenerate interaction of the same kind as the nondegenerate ones. (e): nonideal concurrence due to a ``leaking" interaction. In (c,d,e), entanglement is limited to a number of modes much smaller than $N$.  }
\vspace{-.3cm}
\label{graph}
\end{figure}

It is worthwhile to discuss the relation of the multipartite CV entangler studied in this paper to the multipartite CV entangler proposed by van Loock and Braunstein \cite{vanLoock2000} and realized recently \cite{Aoki2003}. The latter is comprised of $N$ squeezers (some of which may be replaced by vacuum inputs if non-maximal entanglement is acceptable) and an $N$-input-port interferometer. The experiment therefore requires stabilization of the frequencies and output powers of $N$ optical parametric oscillators (OPO), as well as of the $N$ optical paths of the interferometer. In contrast, our system is based on a single OPO pumped by a frequency comb (which can be easily provided by a single CW laser, frequency- or amplitude- modulated, or by a single, stable, pulsed laser) and there is no interferometer at the output. However, the nonlinear medium inside the OPO must provide what we have named concurrent nonlinear interactions, i.e.\ simultaneously phase-matched three-field mixings where each field produced comes from two of the nonlinear processes. (Fig.~\ref{graph}).

For example, the following Hamiltonian
\begin{equation}
\hat H = i\hbar\left(\chi_a\hat b_{12}\hat a_1^\dag\hat a_2^\dag +\chi_b\hat b_{23}\hat a_2^\dag\hat a_3^\dag\right) + H.c.,
\label{2conc}
\end{equation}
where the $\hat b_{ij}$ denote the pump fields, describes a concurrence, even though the multipartite entanglement available from this interaction is not very strong~\cite{Pfister2004,nosso2005}. If $\hat a_2^\dag$ and $\hat a_2$ are swapped in the second term \cite{Ferraro2004,Chirkin2004,Gao2005}, the available tripartite entanglement increases a little, although it is still not as strong as for the system we analyse in the present work. In the graphical representation of Fig.~\ref{graph}, the case of Eq.~(\ref{2conc}) corresponds to three vertices connected by only two edges, i.e.\ an incomplete triangle, analog to Fig.~\ref{graph}(c). 

In the general case, the following concurrence generates a set of $N$ entangled modes $a_{1,\dots,N}$ \cite{Pfister2004}:
\begin{equation}
\hat H_{(N)} = i\hbar\chi\sum_{i=1}^N\sum_{j > i} \hat b_{ij} \hat a_i^\dag\hat a_j^\dag + H.c.
\label{Nconc}
\end{equation}
where we have taken all nonlinear coupling constants equal to $\chi$ for simplicity. The pump fields $\hat b_{ij}$ were assumed to be classical undepleted fields in Ref.~\cite{Pfister2004}, whereas they will be quantized in the present analysis. The analysis of Refs.~\cite{Pfister2004,unpub} shows that the Hamiltonian $\hat H_{(N)}$ leads to perfect multipartite entanglement and corresponds to Fig.~\ref{graph}(a). However, one must: {\em (i)} have all possible pair couplings, {\em (ii)} avoid all couplings outside the set of $N$ entangled modes, and {\em (iii)} carefully suppress all degenerate interactions (i.e.\ terms of the form $\hat b_{i+k,i-k}\ \hat a_i^{2\dag}$) since just a few of these destroy entanglement very quickly. Examples of such undesirable configurations are given in Fig.~\ref{graph}(c,d,e). Their quantitative effects are presently under study and the results will be published elsewhere \cite{unpub}. 

It is important to note that degenerate interactions do not destroy entanglement if they have the opposite effect from the nondegenerate ones, i.e.\ upconverting if the nondegenerate ones are downconverting, or vice versa. This is expressed in the Hamiltonian by opposite signs for the nondegenerate and degenerate terms \cite{Pfister2004} and, in Fig.~\ref{graph}, by solid and dashed lines, respectively. In this case, if all possible degenerate interactions are present with equal coupling strength [Fig.~\ref{graph}(b)], then the situation becomes exactly that of the van Loock and Braunstein proposal \cite{Pfister2004}. This is, however, clearly impossible to implement in a single OPO and all degenerate interactions must therefore be suppressed in our proposal. 

In the absence of any degenerate interaction, the relative signs between nondegenerate Hamiltonian terms do not necessarily decrease the amount of entanglement, even though they do change the particular entangled state that is created. For example, for $N$=3, The Hamiltonian $\hat H_{(3)}$ (\ref{Nconc}), with classical and equal pumps, is
\begin{equation}
\hat H_{(3)} = i\hbar\chi\left[\hat a_1^\dag\hat a_2^\dag+ \hat a_2^\dag\hat a_3^\dag+ \hat a_3^\dag\hat a_1^\dag\right] + H.c.
\end{equation}
and admits eigenstates of the following (unnormalized) form, in the amplitude quadrature basis (and to constant amplitude shifts left)
\begin{equation}
|\psi_{(3)}\rangle = \int | x\rangle_1 |x\rangle_2 |x \rangle_3\ dx
\end{equation}
where $p_o$ is a constant parameter. When one sign is changed in $\hat H_{(3)}$, for example, 
\begin{equation}
\hat H'_{(3)} = i\hbar\chi\left[-\hat a_1^\dag\hat a_2^\dag+ \hat a_2^\dag\hat a_3^\dag+ \hat a_3^\dag\hat a_1^\dag\right] + H.c.,
\end{equation}
the eigenstate becomes of the form
\begin{equation}
|\psi'_{(3)}\rangle = \int | p\rangle_1 |p\rangle_2 |-p \rangle_3\ dp.
\end{equation}
which is a state orthogonal to $\psi_{(3)}$ but is still maximally entangled. Another example is
\begin{eqnarray}
\hat H''_{(3)} &=& i\hbar\chi\left[-\hat a_1^\dag\hat a_2^\dag- \hat a_2^\dag\hat a_3^\dag+ \hat a_3^\dag\hat a_1^\dag\right] + H.c. \\
|\psi''_{(3)}\rangle &=& \int | x\rangle_1 |-x\rangle_2 |x\rangle_3\ dx.
\end{eqnarray}
It is thus easy to see that no entanglement is lost with sign changes for nondegenerate interactions between three modes. It is reasonable to assume it might still be the case for $N>3$ but we haven't yet investigated this more general situation.

Finally, an important fundamental point is that, as recently shown by Braunstein \cite{Braunstein2005}, there exists a formal mathematical connection between our entangler and that of van Loock and Braunstein: a linear algebraic transformation called the Bloch-Messiah reduction projects our system onto theirs. (In technical terms, the Bloch-Messiah reduction is a particular case of singular-value decomposition of the Bogoliubov transformation matrix of the system.)

The ideal concurrences of Fig.~\ref{graph}(a)  would be extremely difficult to obtain using birefringently phase-matched nonlinear optics; however, the advent of quasi-phase-matching has radically changed this situation, as explained in Ref.~\cite{Pfister2004},  and an experimental demonstration of three simultaneous nonlinear interactions, involving the same set of wavelengths in the same nonlinear crystal, has recently been achieved~\cite{Pooser2005}. A concrete and natural way of implementing $\hat H_{(N)}$ experimentally is to have the entangled modes' labels $i,j$ correspond to the eigenfrequencies of an optical resonator, such as the resonant cavity of an  OPO. In that case, additional mode quantum numbers are provided by the optical polarization, a two-dimensional eigenspace. It is then possible to experimentally design quasi-phase-matched concurrences for closed sets of modes, up to $N=4$, using setups such as the one described below. (See Refs.~\cite{Pfister2004,Pooser2005} for the experimental details.) These setups are free of any additional terms deleterious to entanglement. In order to achieve scaling of entanglement to a larger number $N>4$ of modes \cite{idea}, it appears mandatory to use a higher-dimensional quantum number than polarization. The angular momentum of light created in Laguerre-Gaussian beams \cite{LGref} might be an interesting possibility to explore.

\begin{figure}[!htb]
\includegraphics[width=1\columnwidth]{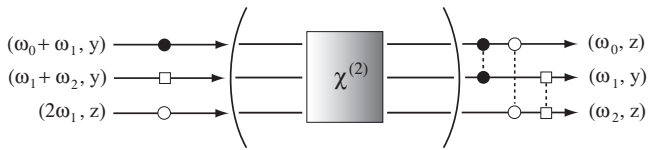}
\caption{Schematic of the experimental setup. Pump lasers drive three modes (denoted by dots, squares and circles), with suitable frequencies and polarisations, which are down converted to three other output modes by the crystal. Note that the physical beams will overlap in the crystal, the separation being intended here for clarity.}
\vspace{-.3cm}
\label{fig:experiment}
\end{figure}

We now describe the experimental system analyzed in detail in this article. A schematic of our system is given in Fig.~\ref{fig:experiment}, showing the three quasi phase-matched pump inputs, which interact with the crystal to produce three output beams at frequencies $\omega_{0}$, $\omega_{1}$ and $\omega_{2}$, with the interactions selected to couple distinct polarisations.
Mode 1 is pumped at frequency and polarisation $(\omega_{0}+\omega_{1},y)$ to produce modes 4 $(\omega_{0},z)$ and 5 $(\omega_{1},y)$, mode 2 is pumped at $(\omega_{1}+\omega_{2},y)$ to produce modes 5 and 6 $(\omega_{2},z)$, while mode 3 is pumped at $(2\omega_{1},z)$ to produce modes 6 and 4. Note that $x$ is the axis of propagation within the crystal. 
It is important to note that no other interaction involving the three output modes is phase-matched in this system. To briefly confirm the feasibility of this scheme: first, it is easy to fulfill the cavity resonance conditions for such a {\em closed} set of 3 coupled modes, the problem being similar to the well-known doubly resonant type-II OPO. Second, type-I cluster instabilities, which are expected above threshold for the $zzz$ interaction, can be avoided by making the frequencies different enough for dispersion to be significant in the nonlinear medium, or by adding a mode filter inside the cavity. We will now turn to the full quantum analysis of the concurrence-based OPO, below and above threshold.

The Hamiltonian for the six-mode system  is
\begin{equation}
\hat{H}_{\rm tot}=\hat{H}_{\rm free}+\hat{H}_{\rm pump}+\hat{H}_{\rm int}+\hat{H}_{\rm damp},
\end{equation}
where the rotating frame interaction Hamiltonian is
\begin{equation}\label{eq:Hint}
\hat{H}_{\rm int} = i\hbar(\chi_{1}\hat{a}_1\hat{a}_4^\dag \hat{a}_5^\dag+\chi_{2}\hat{a}_2\hat{a}_5^\dag \hat{a}_6^\dag+\chi_{3}\hat{a}_3\hat{a}_6^\dag \hat{a}_4^\dag)+{\rm H.c.},
\label{eq:ham}
\end{equation}
with the $\chi_{j}$ representing the effective nonlinear couplings, and the free and damping Hamiltonians have their usual forms. The external pump fields (but not the intracavity ones) are treated as classical. Note that the analysis of Ref.~\cite{Pfister2004}, in contrast, was limited to $\hat{H}_{\rm int}$ with all pump modes being classical and no cavity whatsoever. Its conclusions were therefore applicable only to the optical parametric amplifier case, not to the OPO above threshold, one of the situations treated here.
Following the usual route~\cite{QO} we obtain the master equation
\begin{equation}
\dfdx{\hat{\rho}}{t}=-\frac{i}{\hbar}[\hat{H}_{\rm pump}+\hat{H}_{\rm int},\hat{\rho}]+\sum_{j=1}^6\gamma_j{\cal D}_j[\hat{\rho}]
\label{eq:meq}
\end{equation}
where the Lindblad superoperator ${\cal D}_j[\hat{\rho}]\equiv 2\hat{a}_j\hat{\rho}\hat{a}_j^\dag-\hat{a}_j^\dag\hat{a}_j\hat{\rho}-\hat{\rho}\hat{a}_j^\dag\hat{a}_j$
is obtained by tracing over the density matrix for the usual zero-temperature Markovian reservoirs. We assume that all intracavity modes are resonant with the cavity, although detuning can easily be added.
We will treat all of the high (low) frequency modes as having the same cavity loss rate $\gamma$ ($\kappa$). In what follows we set
$\gamma_1=\gamma_2=\gamma_3=\gamma$, and 
$\gamma_4=\gamma_5=\gamma_6=\kappa$,
noting that the $\chi^{(2)}$ down conversion efficiencies are not necessarily identical~\cite{Pfister2004}.
\par
We now map the master equation onto a Fokker-Planck equation for the positive-P function~\cite{Drummond1980}. Note that we must use the doubled phase space of the positive-P in order to ensure positive semi-definite diffusion.
We can then establish a correspondence between stochastic amplitudes $\alpha_j$ (and $\alpha_j^+$) and mode operators $\hat{a}_j$ (and $\hat{a}_j^\dag$) respectively, $\alpha_j$ and $\alpha_j^+$ being independent complex variables.
The full quantum dynamics of the system can now be found by solving the It\^o stochastic differential equations
\begin{eqnarray}
d\alpha_1 &=& (-\gamma\alpha_1 + E_1 - \chi_1\alpha_4\alpha_5)\hspace{.05cm}dt,\nonumber\\
d\alpha_2 &=& (-\gamma\alpha_2 + E_2 - \chi_2\alpha_5\alpha_6)\hspace{.05cm}dt,\nonumber\\
d\alpha_3 &=& (-\gamma\alpha_3 + E_3 - \chi_3\alpha_4\alpha_6)\hspace{.05cm}dt,\nonumber\\
d\alpha_4 &=& (-\kappa\alpha_4 + \chi_1\alpha_1\alpha_5^+ + \chi_3\alpha_3\alpha_6^+)\hspace{.05cm}dt\nonumber\\
& & + \sqrt{\chi_1\alpha_1}\;dW_1(t) + \sqrt{\chi_3\alpha_3}\;dW_3(t),\nonumber\\
d\alpha_5 &=& (-\kappa\alpha_5 + \chi_1\alpha_1\alpha_4^+ + \chi_2\alpha_2\alpha_6^+)\hspace{.05cm}dt\nonumber\\
& & + \sqrt{\chi_2\alpha_2}\;dW_2(t) + \sqrt{\chi_1\alpha_1}\;dW_1^*(t),\nonumber\\
d\alpha_6 &=& (-\kappa\alpha_6 + \chi_2\alpha_2\alpha_5^+ + \chi_3\alpha_3\alpha_4^+)\hspace{.05cm}dt\nonumber\\
& & + \sqrt{\chi_2\alpha_2}\;dW_2^*(t) + \sqrt{\chi_3\alpha_3}\;dW_3^*(t),
\label{eq:sde6b}
\end{eqnarray}
and also the equations obtained by interchange of $\alpha_j$ with $\alpha_j^+$, and $dW_j(t)$ with $dW_{j+3}(t)$.
The six independent Gaussian complex noises are completely determined by their non-vanishing correlator
\begin{equation}
\overline{dW_i^*(t)dW_j(t^\prime)}=\delta_{ij}\delta(t-t^\prime)\hspace{.1cm}dt.
\label{eq:Wiener}
\end{equation}
Defining the vector $\bm{\alpha}=(\alpha_1,\alpha_1^+,\alpha_2,\alpha_2^+,\dots,\alpha_6,\alpha_6^+)^T$, we determine the stability of the system by linearising the equations about their semi-classical steady state solutions (denoted by $\bar{\bm{\alpha}}$).  We first drop the noise terms in Eq.~\ref{eq:sde6b}, so that $\alpha_j^+\to\alpha_j^*$ and solve for the steady state solutions. The linearised equations for the fluctuations about the steady state 
$\bm{\delta\alpha}=\bm{\alpha}-\bar{\bm{\alpha}}$ then form a multivariate Ornstein-Uhlenbeck process~\cite{QN}. The resulting matrix equation is 
\begin{equation}
d\hspace{.05cm}\bm{\delta\alpha}=\bar{\bm{A}}\bm{\delta\alpha}\hspace{.05cm}dt+\bar{\bm{B}}\hspace{.05cm}\bm{dW},
\label{eq:fluctuationslinear}
\end{equation} 
where $\bm{dW}$ is a vector of independent real noises, $\bar{\bm{B}}$ is the noise matrix of Eq.~\ref{eq:sde6b} with the steady-state values inserted, and 
\begin{equation}
\bar{\bm{A}}=\left[\begin{array}{cc}
\bm{A}_{1} & -\bm{A}_{2}\\
(\bm{A}_{2}^{\ast})^{\rm T} & \bm{A}_{3}\end{array}\right],
\label{eq:fluctdriftmat}
\end{equation}
where $\bm{A}_{1}=-\gamma I_{6}$,
\begin{equation}
\bm{A}_{2}=\left[\begin{array}{cccccc}
\chi_{1}\bar{\alpha}_{5} & 0 & \chi_{1}\bar{\alpha}_{4} & 0 & 0 & 0 \\
0 & \chi_{1}\bar{\alpha}_{5}^{\ast} & 0 & \chi_{1}\bar{\alpha}_{4}^{\ast} & 0 & 0 \\
0 & 0 & \chi_{2}\bar{\alpha}_{6} & 0 & \chi_{2}\bar{\alpha}_{5} & 0 \\
0 & 0 & 0 & \chi_{2}\bar{\alpha}_{6}^{\ast} & 0 & \chi_{2}\bar{\alpha}_{5}^{\ast} \\
\chi_{3}\bar{\alpha}_{6} & 0 & 0 & 0 & \chi_{3}\bar{\alpha}_{4} & 0 \\
0 & \chi_{3}\bar{\alpha}_{6}^{\ast} & 0 & 0 & 0 & \chi_{3}\bar{\alpha}_{4}^{\ast}
\end{array}\right],
\end{equation}
and
\begin{equation}
\bm{A}_{3}=\left[\begin{array}{cccccc}
-\kappa & 0 & 0 & \chi_{1}\bar{\alpha}_{1} & 0 & \chi_{3}\bar{\alpha}_{3} \\
0 & -\kappa & \chi_{1}\bar{\alpha}_{1}^{\ast} & 0 & \chi_{3}\bar{\alpha}_{3}^{\ast} & 0 \\
0 & \chi_{1}\bar{\alpha}_{1} & -\kappa & 0 & 0 & \chi_{2}\bar{\alpha}_{2} \\
\chi_{1}\bar{\alpha}_{1}^{\ast} & 0 & 0 & -\kappa &  \chi_{2}\bar{\alpha}_{2}^{\ast} & 0 \\
0 & \chi_{3}\bar{\alpha}_{3} & 0 & \chi_{2}\bar{\alpha}_{2} & -\kappa & 0 \\
\chi_{3}\bar{\alpha}_{3}^{\ast} & 0 & \chi_{2}\bar{\alpha}_{2}^{\ast} & 0 & 0 & -\kappa
\end{array}\right].
\end{equation}
For simplicity in what follows, we will assume that all the nonlinearities are equal, i.e. $\chi_{j}\equiv\chi$. This can be achieved via quasi-phase-matched interactions in 
periodically poled ferroelectrics~\cite{Pfister2004}, and means that we may set the effective coupling rates equal by equating the pump field amplitudes, $E_{j}\equiv E$.

If we treat all the pump fields as real, all the signal fields are also real since the fields must satisfy the phase condition
$\phi_4+\phi_5=\phi_4+\phi_6=\phi_5+\phi_6=0$.
As with the standard optical parametric oscillator there is an oscillation threshold, which for this system occurs at the pump field strength
\begin{equation}\label{stabthresh}
E^{\rm th}=\gamma\kappa/2\chi,
\end{equation}
below which the stable solutions are 
\begin{equation}\label{eq:alphabelowthresh}
\bar{\alpha}_{(1,2,3)} = E/\gamma, \hspace{1cm}\bar{\alpha}_{(4,5,6)} =0.
\end{equation}
At the threshold the fluctuations are not damped and therefore a linear fluctuation analysis is not valid.
If the pumping is increased further, modes 4, 5 and 6 become macroscopically occupied, while the high frequency modes saturate at their threshold value. There is also a  relationship between the driving fields that must be satisfied to have a steady state, which in the case of equal nonlinearities, means 
that all the pump amplitudes must be equal and hence all the signal amplitudes are equal~\cite{sign}.
The resulting steady state solutions
\begin{equation}\label{eq:alphaabovethresh}
\bar{\alpha}_{(1,2,3)}=\kappa/2\chi,\hspace{1cm}\bar{\alpha}_{(4,5,6)} = \sqrt{(E-E^{\rm th})/\chi},
\end{equation}
enable us to investigate the quantum correlations of the system by studying fluctuations around the steady state, from which we will also find the measurable extra-cavity fluctuation spectra~\cite{QN}. 
\begin{figure}[!htb]
\begin{psfrags}%
\psfragscanon%
%
\psfrag{s03}[t][t]{\setlength{\tabcolsep}{0pt}\begin{tabular}{c}$\omega$ (units of $\kappa$)\end{tabular}}%
\psfrag{s04}[b][b]{\setlength{\tabcolsep}{0pt}\begin{tabular}{c}$I^{out}(\omega)$\end{tabular}}%
%
\psfrag{x01}[t][t]{0}%
\psfrag{x02}[t][t]{0.1}%
\psfrag{x03}[t][t]{0.2}%
\psfrag{x04}[t][t]{0.3}%
\psfrag{x05}[t][t]{0.4}%
\psfrag{x06}[t][t]{0.5}%
\psfrag{x07}[t][t]{0.6}%
\psfrag{x08}[t][t]{0.7}%
\psfrag{x09}[t][t]{0.8}%
\psfrag{x10}[t][t]{0.9}%
\psfrag{x11}[t][t]{1}%
\psfrag{x12}[t][t]{-6}%
\psfrag{x13}[t][t]{-4}%
\psfrag{x14}[t][t]{-2}%
\psfrag{x15}[t][t]{0}%
\psfrag{x16}[t][t]{2}%
\psfrag{x17}[t][t]{4}%
\psfrag{x18}[t][t]{6}%
%
\psfrag{v01}[r][r]{0}%
\psfrag{v02}[r][r]{0.1}%
\psfrag{v03}[r][r]{0.2}%
\psfrag{v04}[r][r]{0.3}%
\psfrag{v05}[r][r]{0.4}%
\psfrag{v06}[r][r]{0.5}%
\psfrag{v07}[r][r]{0.6}%
\psfrag{v08}[r][r]{0.7}%
\psfrag{v09}[r][r]{0.8}%
\psfrag{v10}[r][r]{0.9}%
\psfrag{v11}[r][r]{1}%
\psfrag{v12}[r][r]{0}%
\psfrag{v13}[r][r]{1}%
\psfrag{v14}[r][r]{2}%
\psfrag{v15}[r][r]{3}%
\psfrag{v16}[r][r]{4}%
\psfrag{v17}[r][r]{5}%
%
\includegraphics[width=8cm]{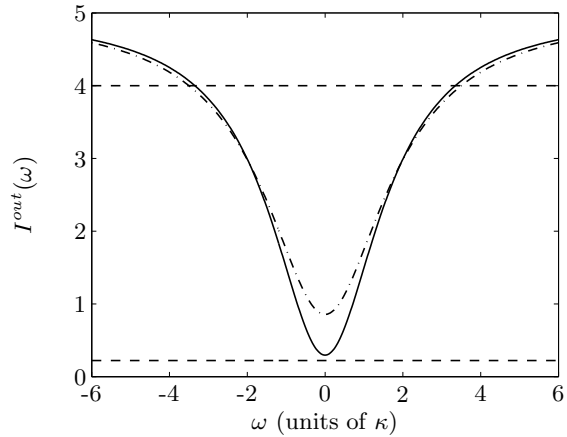}%
\end{psfrags}%
\caption{Spectra for the entanglement criteria of Eq.~\ref{eq:tripart}, for equal pump amplitudes and nonlinearities in each mode. 
In all figures $\chi=0.01$, $\kappa=1$, $\gamma=10$ and  the pump values are $E=0.9E^{\rm th}$ (solid line) and $E=1.1E^{\rm th}$ (dash-dotted line). All quantities plotted in this article are dimensionless. The dashed lines represent the limiting zero-frequency threshold value of $2/9$ and the tripartite correlation value of $4$, which is obviously violated.}
\label{fig:entangle}
\vspace{-.3cm}
\end{figure}
\par
We will first describe the measurable quantities which may be used to experimentally verify that this system exhibits true {\em multipartite} entanglement. 
We define 
quadrature operators for each mode as
\begin{equation}
\hat{X}_j = \hat{a}_j+\hat{a}_j^\dag,\:\:\:
\hat{Y}_j = -i(\hat{a}_j-\hat{a}_j^\dag),
\end{equation}
so that $[\hat{X}_j,\hat{Y}_j]=2i$ and the Heisenberg uncertainty principle requires $V(X_{j})V(Y_{j})\geq 1$, which sets the vacuum noise level. 
Conditions which are sufficient to demonstrate bipartite entanglement for Gaussian variables are well known~\cite{Duan2000}. These have been generalised to tripartite entanglement by Van Loock and Furusawa~\cite{vanLoock2003}, without making any assumptions about Gaussian statistics. Using our quadrature definitions, these conditions give a set of 3 inequalities
\begin{equation}
V(\hat{X}_i-\hat{X}_{j\neq i})+V(\hat{Y}_4+\hat{Y}_5+\hat{Y}_6) \geq 4,
\label{eq:tripart}
\end{equation}
where $i,j \in\{4,5,6\}$ and $V(\hat{A})\equiv\langle \hat{A}^2\rangle-\langle \hat{A}\rangle^2$.
The simultaneous violation of any two of these conditions proves genuine tripartite entanglement for the system. An undepleted pump analysis of the interaction Hamiltonian (Eq.~\ref{eq:Hint}) reveals that these are the maximally squeezed quadrature combinations, so that we can expect optimal entanglement for these variables~\cite{Pfister2004}.
\par
\begin{figure}[!htb]
\begin{psfrags}%
\psfragscanon%
%
\psfrag{s13}[t][t]{\setlength{\tabcolsep}{0pt}\begin{tabular}{c}$\omega$ (units of $\kappa$)\end{tabular}}%
\psfrag{s14}[b][b]{\setlength{\tabcolsep}{0pt}\begin{tabular}{c}$S^{\it inf}(\hat{X}_5 + \hat{X}_6)S^{\it inf}(\hat{Y}_5 + \hat{Y}_6)$\end{tabular}}%
\psfrag{s16}[t][t]{\setlength{\tabcolsep}{0pt}\begin{tabular}{c}$S^{\it inf}(\hat{X}_4)S^{\it inf}(\hat{Y}_4)$\end{tabular}}%
%
\psfrag{x01}[t][t]{0}%
\psfrag{x02}[t][t]{0.1}%
\psfrag{x03}[t][t]{0.2}%
\psfrag{x04}[t][t]{0.3}%
\psfrag{x05}[t][t]{0.4}%
\psfrag{x06}[t][t]{0.5}%
\psfrag{x07}[t][t]{0.6}%
\psfrag{x08}[t][t]{0.7}%
\psfrag{x09}[t][t]{0.8}%
\psfrag{x10}[t][t]{0.9}%
\psfrag{x11}[t][t]{1}%
\psfrag{x12}[t][t]{-6}%
\psfrag{x13}[t][t]{-4}%
\psfrag{x14}[t][t]{-2}%
\psfrag{x15}[t][t]{0}%
\psfrag{x16}[t][t]{2}%
\psfrag{x17}[t][t]{4}%
\psfrag{x18}[t][t]{6}%
%
\psfrag{v01}[r][r]{0}%
\psfrag{v02}[r][r]{0.1}%
\psfrag{v03}[r][r]{0.2}%
\psfrag{v04}[r][r]{0.3}%
\psfrag{v05}[r][r]{0.4}%
\psfrag{v06}[r][r]{0.5}%
\psfrag{v07}[r][r]{0.6}%
\psfrag{v08}[r][r]{0.7}%
\psfrag{v09}[r][r]{0.8}%
\psfrag{v10}[r][r]{0.9}%
\psfrag{v11}[r][r]{1}%
\psfrag{v12}[l][l]{0}%
\psfrag{v13}[l][l]{0.5}%
\psfrag{v14}[l][l]{1}%
\psfrag{v15}[l][l]{0}%
\psfrag{v16}[l][l]{0.5}%
\psfrag{v17}[l][l]{1}%
\psfrag{v18}[r][r]{0}%
\psfrag{v19}[r][r]{2}%
\psfrag{v20}[r][r]{4}%
%
\includegraphics[width=8cm]{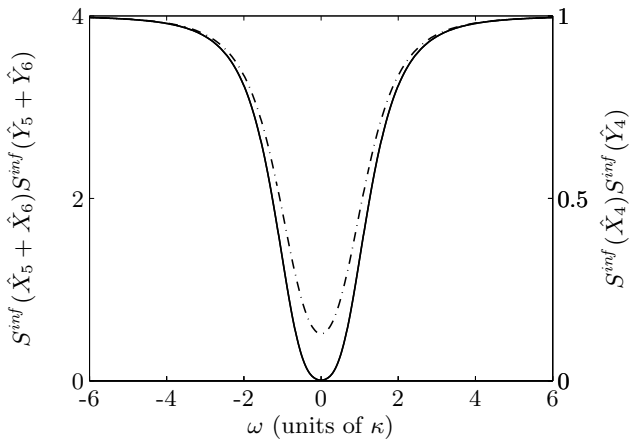}%
\end{psfrags}%
\caption{Output fluctuation spectra of the inferred quadratures for three mode EPR correlations, from measurements on one or two modes. Below threshold ($E=0.9E^{\rm th}$, solid line) the spectra coincide. Above threshold ($E=1.1E^{\rm th}$, dash-dotted line) two mode inference gives better violation than one mode inference, but the difference is indistinguishable on this scale.}
\label{fig:EPR12}
\vspace{-.3cm}
\end{figure}
The Einstein, Podolsky and Rosen (EPR) paradox stems from their 
1935 paper~\cite{Einstein1935}, which showed that local realism
is not consistent with quantum mechanical completeness. Optical quadrature phase amplitudes
have the same mathematical properties as the position and momentum
originally considered by EPR and, as demonstrated by Reid~\cite{Reid1989}, this property can be used to define inferred variances which, when entangled sufficiently, allow for
an inferred violation of the uncertainty principle. This demonstrates the EPR paradox and was experimentally studied by Ou {\em et al.\/}, who found clear
agreement with quantum theory~\cite{Ou1992}.
Following the approach of Ref.~\cite{Reid1989}, we assume that a measurement of the $\hat{X}_{j}$ quadrature, for
example, will allow us to infer, with some error, the value of the
$\hat{X}_{k}$ quadrature, and similarly for the $\hat{Y}_{j,k}$ quadratures.
Minimising this error, we can define variances of the inferred quadratures, the products of which demonstrate the EPR paradox when they appear to violate the Heisenberg uncertainty principle.
\par
Although tripartite entanglement is present in the system, it does not appear that a demonstration of the EPR paradox is possible via the standard two mode approach. There are two ways to demonstrate paradoxes of this nature, both involving information about all three modes. One example is to use quadrature measurements on one mode to infer values for the joint state of the other two. This is mathematically equivalent to inferring, for example, the center of mass position and momenta of two particles by measurements on a third. In this case we find, using mode $i$ to infer properties of the combined mode $j+k$, for example, the inferred variances
\begin{equation}
V^{inf}(\hat{X}_{j}\pm \hat{X}_{k})=V(\hat{X}_{j}\pm \hat{X}_{k})-\frac{\left[V(\hat{X}_{i},\hat{X}_{j}\pm \hat{X}_{k})\right]^{2}}{V(\hat{X}_{i})},
\label{eq:EPR1}
\end{equation}
with similar expressions holding for the $\hat{Y}$ quadratures. In the above $V(\hat{A},\hat{B})=\langle \hat{A}\hat{B}\rangle-\langle \hat{A}\rangle\langle \hat{B}\rangle$ and the $i,j,k$ can be any of $4,5,6$ as long as they are all different. 
The product of the actual variances always satisfies a Heisenberg inequality so that
there is an experimental demonstration of a three mode form of the EPR paradox whenever
\begin{equation}\label{eq:epr1}
V^{inf}(\hat{X}_{j}\pm \hat{X}_{k})V^{inf}(\hat{Y}_{j}\pm \hat{Y}_{k})\geq 4
\label{eq:infer12}
\end{equation}
is violated.
Another option is to make measurements on the combined mode $j+k$ to infer properties of mode $i$. This leads to the inferred variances
\begin{equation}
V^{inf}(\hat{X}_{i})=V(\hat{X}_{i})-\frac{\left[V(\hat{X}_{i},\hat{X}_{j}\pm \hat{X}_{k})\right]^{2}}{V(\hat{X}_{j}\pm \hat{X}_{k})},
\label{eq:EPR2}
\end{equation}
and a demonstration of the paradox whenever
\begin{equation}\label{eq:epr2}
V^{inf}(\hat{X}_{i})V^{inf}(\hat{Y}_{i})\geq 1.
\label{eq:infer21}
\end{equation}
is violated.
\par
The experimentally accessible quantities are the normally ordered fluctuation spectra corresponding to the inequalities defined in Eqs. \ref{eq:tripart}, \ref{eq:epr1}, \ref{eq:EPR2}, for which the same relationships hold. 
These are defined as Fourier transforms of the time-normally ordered operator covariances,
\begin{equation}
S[\hat{A}_i,\hat{B}_j](\omega) \equiv  {\cal FT}\langle:(\hat{A}_i(t)-\langle\hat{A}_i(t)\rangle)(\hat{B}_j(0)-\langle\hat{B}_j(0)\rangle):\rangle,
\label{eq:FTdef}
\end{equation}
which are related to the measurable output spectra, $S^{out}[\hat{A}_i,\hat{B}_j](\omega)$, using the standard input-output relationships for optical cavities~\cite{Gardiner1985}. In the results presented here, we will treat the output mirror as the only source of damping, which is a common approximation in theoretical quantum optics.
For the inequalities of Eq. (\ref{eq:tripart}) we define the fluctuation spectra
\begin{eqnarray}
I^{out}_{ij}(\omega) = S^{out}[\hat{X}_i-\hat{X}_{j\neq i}]+S^{out}[\hat{Y}_4+\hat{Y}_5+\hat{Y}_6],
\label{eq:outspek}
\end{eqnarray}
and use the abbreviation $S^{out}[\hat{A}_j]\equiv S^{out}[\hat{A}_j,\hat{A}_j]$. Similarly, for the variances in Eqs.~\ref{eq:infer12}, \ref{eq:infer21}, we obtain variances of fluctation spectra ($S^{inf}(\omega)$) which must violate the same inequalities in the frequency domain.
\par
We can find relatively simple analytic expressions for the correlations of Eq. \ref{eq:outspek} in the case where all the pump amplitudes and nonlinearities are equal. We denote the sub-threshold solutions (Eq. \ref{eq:alphabelowthresh}) by $\alpha\equiv\bar{\alpha}_{(1,2,3)}$, and the super-threshold solutions (Eq. \ref{eq:alphaabovethresh}) by $\beta\equiv\bar{\alpha}_{(4,5,6)}$. The symmetry means that $I_{45}^{out}(\omega)=I_{46}^{out}(\omega)=I_{56}^{out}(\omega)\equiv I^{out}_{\pm}(\omega)$, where $-(+)$ denotes the spectra below (above) the oscillation threshold. The measurable spectra then take the form
\begin{widetext}
\vspace{-.5cm}
\begin{eqnarray}
\label{eq:below_thresh1}
I_{-}^{out}(\omega)&=& 
5-\frac{8\chi\kappa\alpha[7(\chi\alpha)^2+10\chi\kappa\alpha+4(\omega^2+\kappa^2)]}
{\left[(\chi\alpha+\kappa)^{2}+\omega^{2}\right]\left[(2\chi\alpha+\kappa)^{2}+\omega^{2}\right]},\\
\nonumber\\
I_{+}^{out}(\omega)&=&5-\frac{4\kappa^2(\omega^2+\gamma^2)[76(\chi\beta)^4+(\chi\beta)^2(100\gamma\kappa-56\omega^2)+(\omega^2+\gamma^2)(16\omega^2+43\kappa^2)]}{\left[\left(4(\chi\beta)^2+2\gamma\kappa-\omega^{2}\right)^{2}+\left(2\kappa+\gamma\right)^{2}\omega^{2}\right]
\left[\left(2(\chi\beta)^2+3\gamma\kappa-2\omega^{2}\right)^{2}+\left(3\kappa+2\gamma\right)^{2}\omega^{2}\right]}.
\end{eqnarray}
\end{widetext}
The solutions for these quantities are shown for pump field amplitudes of $0.9E^{\rm th}$ and $1.1E^{\rm th}$ in Fig.~\ref{fig:entangle}. Below threshold the vacuum outputs exhibit near total violation of the van Loock-Furasawa entanglement criteria near zero frequency and approach the uncorrelated limit of 5 for large frequency. Above threshold the outputs form bright beams and maintain a large violation of the inequalities near zero frequency, which implies strong tripartite entanglement. We note that at $E=1.1E^{\rm th}$ the intensities of the three low-frequency modes are twice those of the high frequency modes.  Fig.~\ref{fig:EPR12} shows strong inferred violations of Eqs.~\ref{eq:epr1},~\ref{eq:epr2} for the same degree of entanglement shown in Fig.~\ref{fig:entangle}, indicating that above threshold the system can be used to demonstrate three mode EPR correlations with bright output beams.
\par
In conclusion, we have shown that concurrent intracavity $\chi^{(2)}$ nonlinearities can be used to produce strongly tripartite entangled outputs, both above and below the oscillation threshold. The above threshold entanglement is found for macroscopic intensities, demonstrating a bright tripartite entanglement resource. We have also shown how this device can be used to perform multi-mode demonstrations of the EPR paradox. We expect that this proposal will be easier to stabilise than devices which rely on separate OPOs and multi-port interferometers.
\par
This research was supported by the Australian Research Council and the National Science Foundation grant Nos. PHY-0245032 and EIA-0323623 and by the NSF IGERT SELIM program at the University of Virginia. We thank Peter Drummond for useful discussions.

\begin{thebibliography}{10}
\bibitem{idea} Our initial aim was, indeed, to create a Schr\"odinger-cat-like state, i.e.\ to entangle a large number of cavity modes by pumping, for example, with the frequency comb of a stable femtosecond laser.
\bibitem{sign}
Due to the presence of the square-root, there is an ambiguity in the sign of these solutions. Closer analysis shows that stable solutions all have the same sign.
\end{thebibliography}

\end{document}